\documentstyle[psfig,twocolumn,aps]{revtex}
\begin{document}

\draft
\title{Topological (Sliced) Doping of a 3D Peierls System:
Predicted Structure of Doped BaBiO$_3$}

\author{ Ilka B. Bischofs\cite{byline1}, Philip B. Allen,
Vladimir N. Kostur\cite{byline2}, Rahul Bhargava\cite{byline3}}
\address{Department of Physics and Astronomy, State University of New York,
Stony Brook, New York 11794-3800}

\maketitle

\begin{abstract}

At hole concentrations below $x$=0.4, Ba$_{1-x}$K$_x$BiO$_3$ is
non-metallic.  At $x=0$, pure BaBiO$_3$ is a Peierls insulator.
Very dilute holes create bipolaronic point defects in the Peierls
order parameter.  Here we find that the Rice-Sneddon version of
Peierls theory predicts that more concentrated holes should form
stacking faults (two-dimensional topological defects, called
slices) in the Peierls order parameter.  However, the long-range
Coulomb interaction, left out of the Rice-Sneddon model,
destabilizes slices in favor of point bipolarons at low
concentrations, leaving a window near 30\% doping where the sliced
state is marginally stable.

\end{abstract}


\section{introduction}
When a 1D crystal is driven incommensurate, for example, by doping
to alter the Fermi wavevector, it is well-known that the
modulation of the commensurate crystalline order tends not to be
uniform but to accumulate at kinks in the crystalline order
parameter \cite{Pokrovsky}. A prototype for this is the
Su-Schrieffer-Heeger model \cite{Su} for polyacetylene, (CH)$_x$.
In a simple chemical view, the carbon chain has alternating single
and double bonds. Crystallography confirms this lattice
dimerization. The scalar order parameter (amplitude of the
staggered charge disproportionation) has two possible values,
$\pm\rho$.  Zero-dimensional domain walls (topological defects)
separate regions of the 1D chain with positive and negative order
parameters.  Su {\it et al.} showed that when carriers are doped
into polyacetylene, they create new domain-wall defects and
localize into mid-gap electronic soliton states on these defects.
Analogous effects in two dimensions \cite{Zaanen} occur in layered
materials near a Mott insulator phase, such as cuprates,
nickelates, and layered manganites.  The one-dimensional domain
wall in the 2D system is generally called a ``stripe.''  The 3D
analog, which occurs in some cubic manganites \cite{Mori}, is a
planar domain wall \cite{Khomskii}, sometimes called a ``sheet''
or ``lamella.''  We prefer the term ``slice,'' which, like the
word ``stripe'' can be used as verb or noun.  Here we discuss a 3D
model system in which bonding and strain energies favor slices,
but long-range Coulomb interactions prefer point defects.  We
analyze the system numerically and show that at doping levels near
30\%, Ba$_{1-x}$K$_x$BiO$_3$ is near the crossover where slices
may become more favorable than point defects for lowering the
energy of the doped-in holes.

Charge-density-wave (CDW) and Peierls systems have scalar order
parameters $\rho_Q=N^{-1}\sum_{\ell}
\rho_{\ell}\exp(i\vec{Q}\cdot\vec{\ell})$.  BaBiO$_3$ is a simple
3D example.  With nominal valence Bi$^{4+}$, the Bi 6$s$-band is
half-filled.   The material is non-metallic, with a 2eV optical
gap \cite{Lobo}.  Crystallography \cite{Cox} shows a doubled unit
cell; $\rho_{\ell}$ alternates with wavevector $Q=(\pi,\pi,\pi)$.
The simple cubic sublattice of nominal Bi$^{4+}$ ions
self-organizes into a bipartite (rocksalt-type) charge-ordered
array of nominal Bi$^{3+}$ ($\rho=4-\rho_Q$, called ``A'' sites)
and nominal Bi$^{5+}$ ($\rho=4+\rho_Q$, called ``B'' sites) ions.
The actual value of the order parameter $\rho_Q$ has magnitude
$\le 1$, and takes two degenerate values, $\pm|\rho_Q|$.  A simple
way to think of this is as a 3D version of a Peierls instability.
In a bi-partite lattice with only nearest-neighbor hopping there
is accidental Fermi surface nesting:
$\epsilon(\vec{k})=-\epsilon(\vec{k}+(\pi,\pi,\pi))$, and both are
zero at the Fermi energy.  This guarantees that the crystal can
reduce its electronic band energy {\it via} the electron-phonon
interaction by dimerizing.  A Hamiltonian which contains this
effect was introduced by Rice and Sneddon \cite{Rice,Jurczek}.

\par
\begin{figure}[t]
\centerline{\psfig{figure=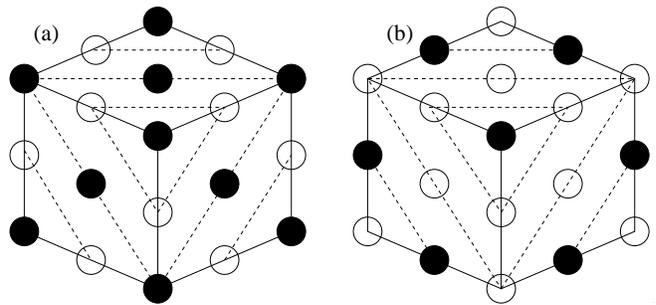,height=1.6in,width=3.4in,angle=0}}
\caption{Fragment of BaBiO$_3$, showing Bi atoms only.  Filled
circles are Bi$^{3+}$ (A) ions, and empty circles are Bi$^{5+}$
(B) ions. In (a) the normal ABABAB... sequence of (111) layers is
shown; in (b), a hole-type stacking fault ABBABA..., or slice, is
shown.}
\label{fig:struc}
\end{figure}
\par

Regions of charge-ordered BaBiO$_3$ with $\rho_Q>0$ are separated
from regions with $\rho_Q<0$ by stacking faults. The simplest
stacking fault lies in a (111) plane.  In perfectly ordered
BaBiO$_3$, (111) planes are alternating A and B type (Fig.
\ref{fig:struc}a).  A stacking fault with no nuclear disorder has
either two adjacent A layers (local charge excess -1 per site on
the plane, or electron-doped) or two adjacent B layers (local
charge excess 1, or hole-doped) as in Fig. \ref{fig:struc}b.

Here we point out that the Rice-Sneddon Hamiltonian, given below,
predicts that holes or electrons, when introduced, will
self-organize into slices.  However, the long-range Coulomb
interaction, neglected in the Rice-Sneddon model, will destabilize
slices except possibly at the most favorable doping level.  For
light doping, the preferred structure for holes is to
self-organize into point defects, small bipolarons, as previously
discussed \cite{Yu,Allen1,Bischofs,Allen2}.

\section{Rice-Sneddon Model}

The Rice-Sneddon model \cite{Rice} is simple, well-studied
\cite{Jurczek,Yu,Iwano,Piekarz}, and quite successful
\cite{Bischofs,Allen2}.  In the perovskite crystal structure of
BaBiO$_3$, each Bi atom is surrounded by six oxygens, and each
oxygen is shared by two Bi atoms.  At high temperatures (of order
1000K) the crystal is nearly cubic perovskite, but at lower $T$
there are rotations and distortions of the BiO$_6$ octahedra,
which enlarge the unit cell.  We believe that the most important
effect is the ``breathing'' displacements $u_0\approx \pm 0.12\AA$
\cite{Cox,Chaillout,Boyce} of oxygens away from or toward
alternate Bi atoms.  These provide a natural electron-phonon
mechanism to enhance the Bi 6$s$ charge density on A sites where
oxygens breath outward, and reduce the Bi 6$s$ charge density on B
sites where the oxygens breathe in.  A microscopic Hamiltonian
containing the minimal necessary electron-phonon interaction was
given by Rice and Sneddon \cite{Rice},
\begin{eqnarray}
{\cal H}=&-&t\sum_{<\ell,\ell'>}c^{\dagger}_{\ell} c^{}_{\ell'}
-g\sum_{\ell}e(\vec{\ell})c^{\dagger}_{\ell} c^{}_{\ell} \nonumber
\\
 &+&\frac{1}{2} K
\sum_{\ell,\alpha}u(\vec{\ell}+\frac{\hat{\alpha}}{2})^2.
\label{eq:ham}
\end{eqnarray}
The first term is nearest-neighbor hopping of Bi $6s$ electrons
with hopping integral $t\approx$ 0.35 eV.  The index of summation
$\ell$ implicitly includes a spin as well as site quantum number.
The filling is $1-x$ electrons per site.  The variable
$u(\vec{\ell}+\hat{\alpha}/2)$ (with $\alpha=x,y,z$) is the
displacement along a Bi-O-Bi bond in the $\hat{\alpha}$ direction
of the oxygen located at position $(\vec{\ell}+\hat{\alpha}/2)a$.
The variable $e(\ell)$ is the local dilation or ``breathing''
amplitude of the 6 oxygens which surround the Bi ion at site
$\ell$,
\begin{eqnarray}
e(\ell)&=&[u(\vec{\ell},x+\hat{x}/2)-u(\vec{\ell}-\hat{x}/2)] +
[x\rightarrow y]  \nonumber \\
 &+& [x\rightarrow z]. \label{eq:dil}
\end{eqnarray}
The Einstein restoring force $K\approx$ 19 eV/$\AA^2$ is fitted to
the measured 70 meV frequency of the Raman-active Peierls
breathing mode \cite{Tajima}.

At half-filling, this model opens a Peierls gap at the Fermi
level; the electron-phonon interaction parameter $g\approx$ 1.39
eV/$\AA$ is fitted to the measured \cite{Lobo} $\approx$2 eV gap.
This is an ordinary size electron-phonon coupling.  The change in
Coulomb field of a charge -2$e$ oxygen ion gives $g\approx$ 1.2
eV/$\AA$. The resulting dimensionless coupling constant
$\Gamma\equiv g^2/Kt$ is $\approx$0.30, intermediate between the
weak ($\Gamma<0.2$) and strong ($\Gamma>0.4$) coupling regimes. In
this middle regime, neither hopping nor electron-phonon energy
dominates \cite{Bischofs}. The ground-state of undoped BaBiO$_3$
is as close to a bipolaronic crystal (large $\Gamma$, $|\rho_Q|
\approx 1$) as to the conventional Peierls-CDW (small $\Gamma$).
We calculated \cite{Bischofs} the order parameter $\rho_Q$ at
$\Gamma=0.3$ to be 0.82.  The corresponding oxygen displacement
$u_0=2g\rho_Q /K=0.12\AA$ agrees with the diffraction and EXAFS
measurements \cite{Cox,Chaillout,Boyce}, showing that the model is
internally consistent.

We previously found that excitations across the Peierls gap form
self-trapped excitons \cite{Allen2}. We also reported that holes
inserted into BaBiO$_3$ self-trap and form bipolarons
\cite{Bischofs} since the coupling strength exceeds
$\Gamma_c=0.17$. These are doubly charged point defects,
corresponding to local depressions of the order parameter where
the oxygen distortion $e(\vec \ell) \rightarrow 0$ for $t=0$. For
non-zero hopping $t$ the bipolaron spreads out and evolves
continuously from a small bipolaron ($\Gamma \gg \Gamma_c$) to
large CDW-like bipolaron as $\Gamma \rightarrow \Gamma_c$. The
stability of bipolaron defects provides a simple explanation why
dilutely doped BaBiO$_3$ remains insulating and diamagnetic.

The ``disproportionation reaction''
2Bi$^{4+}\rightarrow$Bi$^{3+}$+Bi$^{5+}$ has been much discussed
in the literature on BaBiO$_3$. Using the definition
$2U=E($Bi$^{3+})+E($Bi$^{5+})-2E($Bi$^{4+})$, one can say that the
effective  Hubbard $U$ parameter is negative.  Two factors may
contribute to $U$, electron-phonon effects (expected to be
attractive) and Coulomb repulsion ($U=U_{\rm ep} +U_{\rm el}$). If
one wants to assign the mechanism for disproportionation
completely to Coulomb interactions, then the Hubbard $U$
calculated with all atoms held stationary in cubic perovskite
positions (defined as $U_{\rm el}$) should be negative. Vielsack
and Weber \cite{Vielsack} did careful calculations of $U_{\rm
el}$, finding no evidence for negative values, but instead a small
positive value $U_{\rm el}\approx 0.6 \pm 0.4$eV.  Therefore we
shall temporarily ignore the on-site Coulomb repulsion $U_{\rm
el}$.  Apparently the Bi$^{4+}$ ion must reorganize its
environment in order to stabilize the charge disproportionation.

\section{Hole doping}

What happens at finite doping concentrations $x$? Assuming
sufficient electron-phonon coupling to destroy the undistorted
metal, there are two possibilities, (1) bipolarons, and (2)
slices.

\begin{figure}[t]
\centerline{\psfig{figure=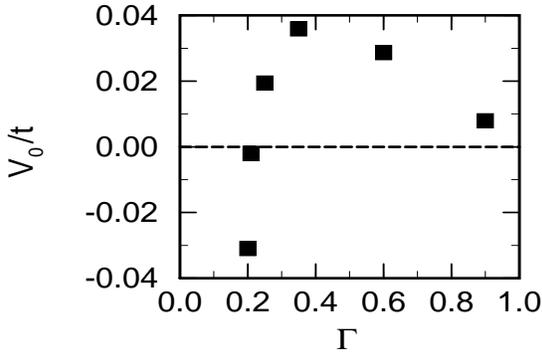,height=2.0in,width=3.in,angle=0}}
\caption{Interaction potential $V_0$ for two bipolarons sitting on
$(0,0,0)$ and $(a/2)(1,1,0)$. Large bipolarons attract whereas
small bipolarons repel.} \label{fig:2}
\end{figure}

(1) Numerical studies by Yu {\it et al.} \cite{Yu}, and confirmed
by us, show that randomly located point bipolarons are at least
metastable.  Depending on whether bipolarons attract or repel, the
system could then either phase-separate into undoped and doped
regions, or form spatially separated bipolarons.  When bipolarons
are small, the energy of an array of bipolarons is described by a
pair-wise additive potential $V(\Gamma, r)$, containing the
repulsive long-range Coulomb interaction $V_{\rm Coul}$ (neglected
for the time being) and the interaction $V_t$, induced by hopping.
$V_t$ decays exponentially with the distance $r$ between two
bipolarons (like bipolaron wave functions.) Since bipolarons can
only sit on former A sites, the nearest neighbor interaction
$V_0=V_t(\Gamma,\sqrt{2}a)$, which is also the strongest
interaction, is between bipolarons separated by $a(110)$.)  We
computed $V_0$ numerically by optimizing the oxygen-positions
self-consistently for given bipolaron positions. Fig. \ref{fig:2}
shows that large CDW-like bipolarons attract each other, whereas
small bipolarons repel, with $V_t\rightarrow0$ in the atomic limit
$t\rightarrow0$.  Perturbation theory around $t=0$ gives a
bipolaron repulsion.  At the physically relevant value $\Gamma
=0.3$, bipolarons repel according to  Fig. \ref{fig:2}, but
multi-bipolaron interactions become important with decreasing
$\Gamma$. Therefore, in the intermediate coupling regime we must
rely on exact numerical diagonalization of Eq. (\ref{eq:ham})
\cite{Yu}.

\par
\par

\begin{figure}[t]
\centerline{\psfig{figure=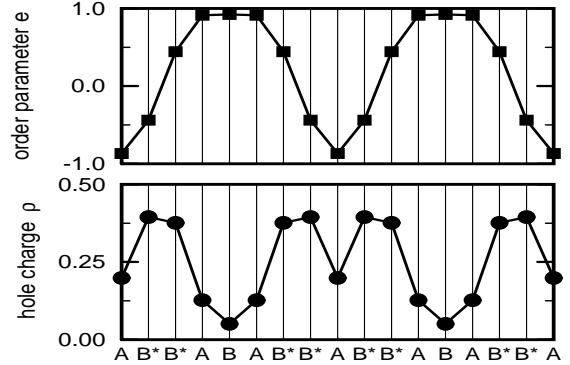,height=2.0in,width=3.in,angle=0}}
\caption{Normalized breathing order parameter $e/6u_0$ (upper panel) and
hole charge $\rho_{\rm hole}$ (lower panel) for a
$x=1/4$-doped phase-slip structure
(AB$^{\ast}$B$^{\ast}$ABAB$^{\ast}$B$^{\ast}$)$_N$ at $\Gamma=0.3$
(two periods shown). The order parameter changes sign across each
phase-slip B$^{\ast}$B$^{\ast}$. Holes are localized on phase-slip
planes with slight spreading to neighboring A-planes.  The hole
charge is computed relative to a reference system with the same
structure, (AB$^{\ast}$B$^{\ast}$ABAB$^{\ast}$B$^{\ast}$)$_N$, but
with A charges fixed at their undoped value 4+$\rho_Q$, B charges
fixed at 4-$\rho_Q$, and B$^{\ast}$ charges fixed at 4-$\rho_Q/2$
corresponding to zero doping, with the charge deficit $\rho_Q$
spread equally on the two adjacent B$^{\ast}$ layers.  The charge
fluctuation $\rho_Q$ has the value 0.82 at $\Gamma=0.3$.}
\label{fig:3}
\end{figure}

(2) In contrast to bipolaron defects where the order parameter
never changes sign, holes could form topological defects.
Phase-slips are planar defects with sign-changes of the charge
order parameter $\rho_Q$ and of the breathing order parameter
$\hat{e}=(-1)^l e_l$.  Consider a ``BB'' stacking fault in the
111-direction (Fig. 1b). In the atomic ($t \rightarrow 0$) limit,
each B-site on a phase-slip plane has three displaced and three
undisplaced oxygen neighbors, i.e. $e_l \approx \pm 3 u_0$.
Phase-slips accumulate one hole for every two atoms on a (111) BB
bilayer.  The average hole charge on phase-slip B-sites is thus
$\rho_{\rm hole}\approx +\frac{1}{2}$. The actual values of hole
charge found for $\Gamma=0.3$ and $x=1/4$ are shown in Fig.
\ref{fig:3}.

\par
\par

Our aim is to find the ground-state hole arrangement, testing the
stability of several bipolaronic (1) versus the phase-slip (2)
solutions numerically.  We did a series of calculations on
$x=1/4$-doped systems.  The bipolaron sytems (1) were (a) maximal
spacing between bipolarons, obtained when they occupy center and
corner sites of a tetragonal 16 atom unit cell ({\it bct}
structure); (b) a simple cubic ({\it sc}) arrangement of
bipolarons sitting at the corner sites of a 8 atom cubic cell; (c)
a disordered structure with random bipolaron positions; (d)
phase-separated structures based on unit cells containing 8 or 16
(111) planes, where we replace one or two near neighbor A planes with
bipolaronic B planes, [(AB)$_3$B$_2$]$_N$ or
[(AB)$_6$B$_4$]$_{N/2}$. We also looked at unit cells with 8 or 16
111 planes containing phase slips (2) [AB(ABB)$_2$]$_N$ and
[(AB)$_2$(ABB)$_4$]$_{N/2}$. Finally, we looked at the undistorted
metal [C]$_N$ where each atom C has a nominal Bi$^{4.25+}$
valence.

Phase separation was strongly disfavored, while separated
bipolarons and phase slips were all metastable.  The phase slips
weakly repelled, preferring the 8 plane solution to the 16 plane
solution. The order parameter and hole charge density of this 8
plane solution are shown in Fig.(\ref{fig:3}). The stability is
determined computing the total energy $E_{\rm tot}(\{u_{\vec
l}\})$ given by Eq.(\ref{eq:ham}), which is a function of the
oxygen-positions $u_{\vec l}$. We start by guessing oxygen
positions to get close to a local minimum in the energy landscape,
then vary oxygen positions using a gradient minimization routine
to find a self-consisistent minimum. For smaller periodic
structure, we used $k$-space sampling in the corresponding
Brillouin zones (8000 k-points). For each $\vec k$, ${\cal H}$ is
diagonalized exactly.  States are filled with two electrons up to
the desired doping. For the random bipolaron structure we used
large asymmetric clusters ($\approx$400 atoms) with periodic
boundary conditions and $k=0$ only. Initial oxygen-positions had
Peierls order with small random deviations. Our calculations on
random configurations generally reproduce the earlier calculations
of Yu {\it et al.} \cite{Yu}.

\par
\begin{figure}[t]
\centerline{\psfig{figure=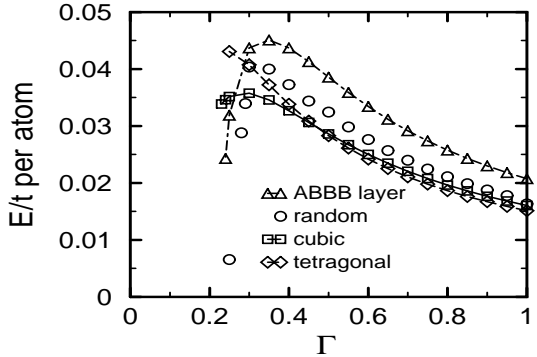,height=2.0in,width=3.in,angle=0}}
\caption{Energy difference $E/t$ of bipolaron structures (1a-d)
relative to the most favorable phase-slip structure (2) for $x=1/4$-doping.}
\label{fig:4}
\end{figure}
\par

Energies of various states at doping $x$=1/4 are shown in
Fig. \ref{fig:4}.
In the atomic limit $t=0$ all bipolaron and phase-slip structures
are degenerate.  All that is required is that no A site should have
an A first neighbor.  Below a critical
coupling strength $\Gamma_c(x)$ ($\approx 0.2$ for doping $x$=1/4),
distorted structures become unstable with respect
to the undistorted metallic structure.  Above $\Gamma_c(x)$,
we find numerically that at $x=1/4$ holes strongly
prefer to order in phase-slips (2).
Bipolaronic structures (1) are quite similar to each other in energy,
and behave as expected from the
the bipolaron-pair interaction. In the intermediate
coupling range there occurs a
cross-over from the tetragonal to the layered BBB-structure as
stable bipolaron configuration,
corresponding to a change in the overall bipolaron interaction from
weakly repulsive to weakly
attractive. For 1/4-doped BaBiO$_3$, the energy gain
for the phase-slip solution compared to the bipolarons is about 50-65 meV
per hole.

Thus, contrary to previous studies of doping of this model
\cite{Jurczek,Yu}, we find that the stable doping state
is not bipolarons but phase slips.  Does this model correspond
sufficiently to reality for BaBiO$_3$?  Iwano and Nasu \cite{Iwano} use a
more complicated hopping and electron-phonon interaction, but we
believe that such corrections are not the relevant ones.  There are also
(i) small structural distortions (rotations of oxygen octahedra) beyond
the breathing-mode distortions considered here, and (ii) non-adiabatic
effects (such as zero point motion) associated with the fact that the
oxygen mass is not infinitely large compared with the electron
mass.   We believe that both of these also have
little relevance.  It is harder to dismiss two other effects:
(iii) the disorder caused by the dopant atoms, and (iv) the long-range
Coulomb interaction, both omitted so far.  Of these, the last is
clearly important, as we now show, and tends to destabilize phase-slips
relative to distributed bipolarons.

\section {Long-range Coulomb effect}

At low doping, there is a large Coulomb cost in putting charges
onto stacking faults instead of widely distributed point charges.
We modeled this as follows.  The Madelung energy was computed by
the Ewald method for
(Ba$^{2+}$)$_2$(Bi$^{3+}$Bi$^{5+}$)(O$^{2-}$)$_6$.  The
calculation was re-done for many large unit cells, with holes
added on selected Bi ions (Bi$^{3+}\rightarrow$Bi$^{5+}$), and
compensating negative charges distributed uniformly throughout
space.  This is one way to mimic potassium doping of BaBiO$^3$.
First consider the sliced state. By numerical calculation for
uniformly distributed phase slips at many values of $x$ between
1/39 and 1/3, we found a good fit to the formula $E_S=(-25.1 +
\pi/9x)( e^2/2a\epsilon_{\infty})$. The static electronic
screening $\epsilon_{\infty}\approx 5$ was measured by Tajima {\it
et al.} \cite{Tajima1}.  $E_S$ is the difference of energy per
hole between the sliced solution and the undoped Peierls
insulator. The term $\pi/9x$ is the analytic result for idealized
uniform sheets of charge of vanishing thickness, arranged
periodically in a compensating charge background.  The term -25.1
corrects for the discreteness of the charges, the absence of the
self-interaction, and includes the Coulomb energy of the holes
with the background BaBiO$_3$ lattice.

\begin{figure}[t]
\centerline{\psfig{figure=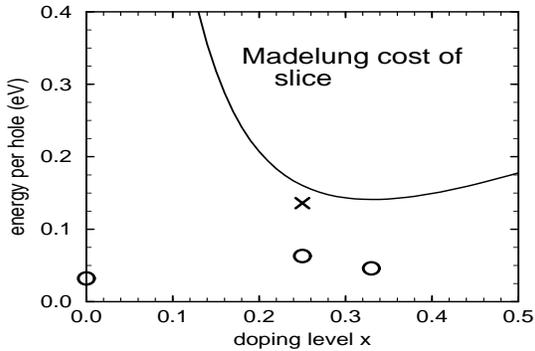,height=2.0in,width=3.in,angle=0}}
\caption{The curve is an analytic approximation to the Madelung or
Coulomb repulsion of a periodic array of (111) phase-slips
(slices) compared with optimally spaced separated bipolarons at
the same average density, where doped-in holes are modelled as
point charges (charge =$e$) with screening $\epsilon_{\infty}$ set
to 5.  The open circles give the numerically computed Peierls
short-range stabilization energy (from hopping and electron-phonon
energy) of sliced solutions relative to separated bipolarons.  The
symbol {\bf x} is the exact numerical Madelung energy difference
at 1/4 doping between the sliced solution and the {\it bct}
bipolaron solution with the numerically computed layer charges
$\rho_i$.} \label{fig:mad}
\end{figure}

We also need the energy difference per hole $E_B$ of optimally
spaced bipolarons relative to the undoped Peierls insulator.  For
doping $x=1/n^3$ the bipolarons can be placed on a sublattice of
face-centered cubic ({\it fcc}) form, with maximum spacing.
Numerical results for $n$=2, 3, and 4 fitted well to the formula
$E_B=(-21.2-4.585x^{1/3})(e^2/2a\epsilon_{\infty})$. The term
$-4.585x^{1/3}$ is the Madelung energy of an {\it fcc} lattice of
charges $2e$ in a uniform compensating background, and the
constant -21.2 accounts for the energy of the holes in the
background BaBiO$_3$ lattice. The difference $E_S-E_B$ is the
Coulomb penalty per hole for forming slices or phase slips.  It is
plotted in Fig \ref{fig:mad}, becoming large at low $x$ with a
shallow minimum near $x=0.3$.

A more accurate estimate of the Coulomb penalty is shown as an
{\bf x} on Fig \ref{fig:mad}, and was obtained by doing Ewald sums
using the actual Bi site charges ($\rho_i=5-n_i$ where
$n_i=2<\psi_i^{\dagger}\psi_i>$ is the local occupancy of the Bi
$s$ states) and the actual
(AB$^{\ast}$B$^{\ast}$ABAB$^{\ast}$B$^{\ast}$)$_N$ slice structure
and {\it bct} bipolaron structure.  It is seen on the figure that
at 1/4 doping, the Coulomb penalty is twice as high as the Peierls
benefit in forming slices.  These calculations did not account for
the actual random positions of the compensating negative charges
where Ba$^{2+}$ ions are replaced by K$^+$ ions.  Instead, the
compensating charge was distributed uniformly. The calculations
were repeated for a model where the compensating negative charges
were equally shared by each Ba atom (Ba$^{2+}\rightarrow$Ba$^{(2-x
)+}$).   This only changed the constant terms in $E_S$ and $E_B$,
but did not affect the difference, $E_S-E_B$ plotted here.

\par
\par

The Madelung sums discussed above are not the complete Coulomb
effect.  The missing on-site repulsion has the opposite effect,
preferring sliced solutions to distributed bipolarons.  If we add
back the local term
\begin{equation}
{\cal H}_U=U_{\rm el}\sum_i (\rho_i-4)^2 \label{eq;U}
\end{equation}
and treat it as a first-order perturbation, the sliced solution at
1/4-doping has lower on-site energy by 0.121$U_{\rm el}$ per hole
than the {\it bct} bipolaron solution.  Using the value $U_{\rm
el} \approx 0.6\pm 0.4$eV from Vielsack and Weber \cite{Vielsack},
the onsite correction $0.07\pm 0.05$eV is potentially sufficient
to re-stabilize the sliced solution.  Thus we can say that the
best place to look for sliced structures in BaBiO$_3$ is in the
range near 1/4 to 1/3 doping, but we cannot predict whether the
sliced solution will be destroyed by disorder or Coulomb effects.

Doping 1/3 is the highest at which a simple sliced solution
($(ABB)_N$) is possible. For higher doping, either the A-planes
acquire polaron defects or the sliced solution is destroyed and
purely bipolaronic states win.

It is worth mentioning that there have been  reports \cite{Pei}
(subsequently attributed to electron beam heating effects
\cite{Verwerft}) of superlattice diffraction spots in doped
BaBiO$_3$, not apparently identical to the superstructures
predicted here. A further search in the doping region $0.2 < x <
0.35$ would be interesting.


\begin{references}

\bibitem[*]{byline1}  Present address: Max-Planck-Institut f\"ur
                        Kolloid- und Grenzfl\"achenforschung,
                        D-14424 Potsdam, Germany.

\bibitem[\dagger]{byline2}  Present address: AIG Financial Products,
                        Westport, CT 06880.

\bibitem[**]{byline3} work done while a student at The Wheatley School,
            11 Bacon Road, Old Westbury, New York 11568;
            present address: Princeton University, Princeton, NJ.


\bibitem{Pokrovsky}     V. L. Pokrovsky and A.L. Talapov,
                        {\it Theory of Incommensurate Crystals},
                        Harwood Academic Publishers, New York, NY,
                        1984.

\bibitem{Su}          W. P. Su, J. R. Schrieffer, and A. J. Heeger,
                        Phys. Rev. B {\bf 22}, 2099 (1980).

\bibitem{Zaanen}        J. Zaanen and O. Gunnarsson,
                        Phys. Rev. B {\bf 40}, 7391 (1989).

\bibitem{Mori}          S. Mori, C. H. Chen, and S.-W. Cheong,
                        Nature {\bf 392}, 473 (2001).

\bibitem{Khomskii}      D. I. Khomskii and K. I. Kugel,
                        Europhys. Lett. {\bf 55}, 208 (2001).

\bibitem{Lobo}          R. P. S. M. Lobo and F. Gervais,
                        Phys. Rev. B {\bf 52}, 13294 (1995).

\bibitem{Cox}           D. E. Cox and A. W. Sleight,
                        Solid State Commun. {\bf 19}, 969 (1976).

\bibitem{Rice}          T. M. Rice and L. Sneddon,
                        Phys. Rev. Letters {\bf 47}, 689 (1982);
                        P. Prelovsek, T. M. Rice, and F. C. Zhang,
                        J. Phys. C {\bf 20}, L229 (1987).

\bibitem{Jurczek}       E. Jurczek and T. M. Rice,
                        Europhys. Lett. {\bf 1}, 225 (1986);
                        E. Jurczek,
                        Phys. Rev. B {\bf 35}, 6997 (1987).

\bibitem{Yu}            J. Yu, X.-Y. Chen, and W. P. Su,
                        Phys. Rev. B {\bf 41}, 344 (1990).

\bibitem{Allen1}        P. B. Allen and V. Kostur,
                        Z. Phys. B {\bf 104}, 605 (1997).

\bibitem{Bischofs}      I. B. Bischofs, V. N. Kostur, and P. B. Allen,
                        Phys. Rev. B {\bf 65}, 115112 (2002).

\bibitem{Allen2}        P. B. Allen and I. B. Bischofs,
                        Phys. Rev. B {\bf 65}, 115113 (2002).

\bibitem{Iwano}     K. Iwano and K. Nasu,
            Phys. Rev. B {\bf 57}, 6957 (1998).

\bibitem{Piekarz}       P. Piekarz and J. Konior,
                        Physica C {\bf 329}, 121 (2000).


\bibitem{Vielsack}      G. Vielsack and W. Weber,
                        Phys. Rev. B {\bf 54}, 6614 (1996).

\bibitem{Chaillout} C. Chaillout and A. Santoro,
            Solid State Commun. {\bf 65}, 363 (1988).

\bibitem{Boyce}     J. B. Boyce, F. G. Bridges, T. Claeson, T. H.
            Geballe, G. G. Li, and A. W. Sleight,
            Phys. Rev. B {\bf 44}, 6961 (1991).

\bibitem{Tajima}        S. Tajima, M. Yoshida, N. Koshizuka, H. Sato,
                        and S. Uchida,
                        Phys. Rev. B {\bf 46}, 1232 (1992).


%

\bibitem{Tajima1}   S. Tajima, S. Uchida, A. Masaki, H. Takagi,
            K. Kitazawa, S. Tanaka, and A. Katsui,
            Phys. Rev. B {\bf 32}, 6302 (1985).

\bibitem{Pei}       S. Pei, N. J. Zaluzec, J. D. Jorgensen,
                    B. Dabrowski, D. G. Hinks, A. W. Mitchell, and D. R. Richards,
                    Phys. Rev. B {\bf 39}, 811 (1989).

\bibitem{Verwerft}  M. Verwerft, G. Van Tendeloo, D. G. Hinks,
                    B. Dabrowski, D. R. Richards, A. W. Mitchell,
                    D. T. Marx, S. Pei, and J. D. Jorgensen,
                    Phys. Rev. B {\bf 44}, 9547 (1991);
                    P. Wochner, Q. J. Wang, S. C. Moss, S. K.
                    Sinha, G. Gr{\"u}bel, H. Chou, L. E. Berman,
                    J. D. Axe, C.-K. Loong, J. Z. Liu, W. D.
                    Mosley, P. Klavins, and R. N. Shelton,
                    Phys. Rev. B {\bf 47}, 9120 (1993);
                    C. H. Du, P. D. Hatton, H. Y. Tang, and M. K.
                    Wu,
                    J. Phys. Cond. Mat. {\bf 6}, L575 (1994).




\end{references}
\end{document}